# Intelligent Approaches to Predictive Analytics in Occupational Health and Safety in India

Ritwik Raj Saxena
Researcher, Department of Computer Science
University of Minnesota.
Duluth, MN, USA – 55812

*Abstract* – Concerns associated with occupational health and safety (OHS) remain critical and often under-addressed aspects of workforce management. This is especially true for high-risk industries such as manufacturing, construction, and mining. Such industries dominate the economic landscape of India which is a developing country with a vast informal sector. Regulatory frameworks have been strengthened over the decades, particularly with regards to bringing the unorganized sector within the purview of law. Traditional approaches to OHS have largely been reactive. Such approaches rely on post-incident analysis (which is curative) rather than preventative intervention. This paper portrays the immense potential of predictive analytics in rejuvenating OHS practices in India. Intelligent predictive analytics is driven by approaches like machine learning and statistical modeling. It improves the process of identification of latent risks, anticipates potential incidents, enhances resource allocation according to better identified needs to prevent workplace hazards before they materialize, and boosts employee trust at work. Its data-driven nature serves to overcome the limitations of conventional OHS methods. Predictive analytics approaches to OHS in India draw on global case studies and generative applications of predictive analytics in OHS which are customized to Indian industrial contexts. This paper attempts to explore in what ways it exhibits the potential to address challenges such as fragmented data ecosystems, resource constraints, and the variability of workplace hazards. The paper also endeavors to highlight barriers to the adoption intelligent predictive in India. It further presents actionable policy recommendations to create conditions conducive to the widespread implementation of this modern approach. Predictive analytics must be advocated as a cornerstone of OHS strategy. In doing so, the paper aims to spark a collaborative dialogue among policymakers, industry leaders, and technologists. It urges a shift towards intelligent practices to safeguard the well-being of India's workforce.

## 1. Introduction

OHS is a multidimensional discipline. In India, a diverse industrial base ranging from traditional manufacturing to rapidly modernizing construction and service sectors presents distinctive challenges in the rejuvenation of the OHS ecosystem. Regulatory frameworks, such as the [now repealed] Factories Act, 1948, and subsequent amendments, have been created. Enforcement mechanisms stay inadequate, and compliance remains superficial. Many organizations continue with incident-driven safety models which put worker health and safety at risk. Such reactive approaches fail to remedy the complex hurdles which arise in the modern occupational landscape. The interplay of a variety of hazards comes into picture. The need for a transformation in how workplace safety is conceptualized and operationalized is pressing.

Predictive analytics employs advanced data mining, statistical modeling, and machine learning algorithms to analyze historical and real-time data. It enables the anticipation of future risks with high precision. Traditional systems rely on retrospective incident analyses and generalized risk assessments. In the manufacturing industry, machinery malfunctions account for the preponderance of workplace injuries. In such industries, predictive maintenance systems have been shown to reduce downtime by a significant percentage. It is also able to mitigate accident risks by preempting equipment failures. In the mining sector, predictive models trained on sensor data from geotechnical monitoring systems are utilized to forecast ground collapses with remarkable accuracy. This saves lives and minimizes economic losses.

India's industrial fabric is poised to benefit significantly from such advancements. This is in light of the proliferation of data collection technologies comprising of IoT-enabled sensors and wearable devices, as well as integrated workplace management systems (IWMS). IoT devices deployed in construction sites continuously monitor worker vitals and environmental parameters. Through a well-crafted, seamless pipeline, this data is fed into predictive models that flag unsafe conditions in real-time. Such systems exemplify the transition from static compliance checklists to resilient safety frameworks that evolve alongside operational needs.

The implementation of predictive analytics in India must account for contextual nuances, including fragmented data ecosystems and resource disparities across industries. The country's informal sector employs over 80% of the workforce. Data collection mechanisms are often absent in this sector. Streamlining the landscape requires innovative solutions such as mobile-based safety applications capable of gathering decentralized data and integrating it into centralized analytics platforms. Integrating predictive analytics with existing enterprise resource planning (ERP) systems enables organizations to align safety metrics with broader operational goals. This ensures a holistic approach to workplace safety.

## 2. Current Challenges and the Need for Predictive Analytics in OHS in India

India's OHS paradigm faces many challenges. The wide spectrum of occupational hazards in India is itself a challenge. Lack of maintenance has caused widespread fraying of electrical wires and exposure of live current-carrying conductors leading to electrocution hazards. When such conductors are exposed to open air, they may come in contact with moisture, especially during monsoons, leading to short circuiting and an even more deleteriously far-reaching hazard of electrocution. Lack of scientificness in the construction of manufacturing facilities has led to an increased risk of physical risks such as falls. Unupdated environmental health and safety (EHS) mechanisms and regulations lead to, among many other risks, the increased risk of exposure to hazardous substances in chemical industries. Unscientific workplace practices also result in ergonomic hazards which may lead to chronic musculoskeletal injuries, driven by long working hours, to unwitting workers. Conditions associated with the human aspects of the workplace may include unchecked authoritarianism by mangers and executives, sexual harassment, intimidation, bullying or other kinds of harassments by colleagues, the mental toll of job insecurity, and inadequate workplace policies for special cases such as pregnancy and parenthood leave for workers who became new parents. Such conditions create psychological hazards in the workplace. These hazards, although less visible, contribute significantly to absenteeism, reduced productivity, and diminished quality of life among workers.

The construction sector employs over 51 million workers in India. In this sector, accidents related to falls from heights remain a predominant risk. Gender disparities in access to personal protective equipment (PPE) results in vastly skewed statistics related to workplace accidents in the construction sector in India where men have proven to have far fewer accidents, incidence of serious workplace injuries (including head injuries and cranial hemorrhages, and fatalities compared to women. In the manufacturing sector, chemical exposure in industries like abrasion industry, leather tanning and textiles is rife. It leads to chronic health conditions such as respiratory disorders like pulmonary sarcoidosis, asbestosis, berylliosis, and Chronic Obstructive Pulmonary Disease (COPD) and skin diseases like chemical acne and squamous cell carcinoma. The mining sector presents risks such as ground subsidence, water inundation, and fire hazards due to methane accumulation. Pneumoconioses like silicosis are also common in the mining sector because of a heightened risk of silica inhalation and exposure to other kinds of particles. The Chasnala incident is a frequently quoted example of a mining disaster in India where more than 370 workers died due to an explosion and subsequent flooding.

One of the primary gaps in India's OHS practices lies in the limited use of data to inform decision-making. Workplace safety audits frequently found using manual checklists that fail to encapsulate and register real-time risks and dynamic workplace conditions. This data vacuum is exacerbated by the lack of standardized reporting mechanisms. This causes safety records to stay in a fragmented and inconsistent state, and such a condition persists across industries. Resource constraints, particularly in small and medium enterprises (SMEs) and the informal sector, hinder the implementation of even basic safety measures. Approximately 90% of India's workforce operates in informal arrangements, a statistic which reduces to 80% in urban backdrops such as New Delhi.

In the informal sector, which is termed so because it is roughly out of reach of the formal regulatory system, safety practices are virtually non-existent. The same goes for the unorganized sector, which largely overlaps with the informal sector and constitutes industries where workers are not organized and unified to have a unanimous voice. This leaves workers vulnerable to preventable injuries and illnesses. In large-scale industries, where safety infrastructure exists, the reliance on conventional risk assessment methods such as risk priority number (RPN) calculations often oversimplifies complex risk scenarios. Traditional metrics fall short of accounting for interconnected factors like environmental variability and human behavior. The outdated nature of these practices is highlighted by the absence of preemptive frameworks. Accident investigations in high-risk industries frequently reveal that warning signs, such as machinery vibrations or abnormal environmental readings, are commonly overlooked and inadequately addressed due to the lack of real-time monitoring and analytics.

Several incidents in recent years underscore the inefficiencies in India's OHS framework. One notable example is the styrene gas leakage from a tank of the firm LG Polymers India Private Limited at Venkatapuram village near Vishakhapatnam, which happened on May 7, 2020. The catastrophe took place when the villagers were peacefully sleeping in their beds, blissfully unaware of what was to come. This early morning cataclysm resulted in 12 fatalities and over 1,000 hospitalizations, including many children. Investigations revealed lapses in monitoring systems and inadequate safety protocols for handling hazardous chemicals. This demonstrates the need for predictive analytics, advanced real-time monitoring, and proactive warning systems in preventing such tragedies.

Another example is the recurring accidents in India's coal mining sector. In December 2016, a wall collapsed at the Lal Matia open cast mine in Jharkhand in India. It led to 17 fatalities, all mine workers. Post-incident analysis showed that early warning signs, including structural instability and unusual seismic activity, were misinterpreted and ignored. The lack of predictive models capable of correlating these signals with collapse probabilities in real-time was felt. In the construction sector, inefficiencies are apparent in the alarming frequency of constructional collapses. On October 23, 2024, at least eight people perished when a building under construction collapsed in Bengaluru in Karnataka in India. It highlighted the absence of sensor-based load monitoring systems which could have triggered preventive measures.

## 3. Role of Predictive Analytics in Enhancing OHS

Predictive analytics has much applicability in OHS. It enables the continuous monitoring of machinery and equipment through IoT-enabled sensors that collect data on temperature, vibration, and wear-and-tear metrics. This data is fed into predictive maintenance models that can preemptively identify machinery likely to fail. This reduces downtime and the likelihood of accidents. Wearable devices equipped with biometric sensors can track the physiological parameters of workers, including their heart rate, pulse rate, breathing rate, sweating, hydration levels, and fatigue. Such arrangements, when equipped with communication and warning systems, can alert supervisors and standby rapid action teams to hazardous conditions that could lead to incidents such as heatstroke and exhaustion. Wearable sensors can also monitor levels of sedentariness, sleeplessness, glucose levels, and so on, and make an employee aware of the risk of non-communicable diseases which may develop due to workplace conditions such as high work stress.

In the construction industry, predictive analytics has been integrated with Building Information Modeling (BIM) systems to forecast structural vulnerabilities and material degradation over time. In mining, predictive geotechnical models analyze seismic activity, subsurface conditions, integrity of the soil and rock structures, gas volume and leakages, and equipment usage data to predict and mitigate the risks of mine collapses and gas explosions. Data from sensors monitoring core temperatures, radiation levels, coolant flow rates, and reactor pressure is fed into machine learning algorithms to predict anomalies such as reactor overheating and coolant system failures. Predictive models use historical inspection data and real-time ultrasonic readings for early detection of stress corrosion cracking in reactor pressure vessels. Such systems minimize risks of radioactive leaks or core meltdowns.

In petroleum mining, predictive analytics integrates seismic data, drilling parameters, and real-time geophysical inputs to forecast blowouts, gas kicks, and equipment failures. Predictive models are employed in offshore drilling platforms to analyze drill string vibrations, mud flow anomalies, and pressure fluctuations to preemptively signal instability in wellbores to enhance worker safety and reduce the environmental risks associated with oil spills. Fractional distillation units in chemical and petrochemical industries operate under extreme thermal and pressure conditions. Resultantly, they are prone to risks such as column flooding and heat exchanger failures. Predictive analytics utilizes data on feedstock properties, temperature gradients, pressure profiles, and so on, to forecast operational anomalies. Advanced models, trained on historical failure data, predict fouling in heat exchangers and structural fatigue in distillation columns. This helps with preemptive maintenance and ensuring process stability. In transportation of oil and gas through pipes, predictive analytics can analyze data on pipeline pressure and corrosion to forecast leaks and ruptures. This diminishes environmental and operational impacts of the petroleum industry.

In physics and chemistry laboratories, experiments may involve volatile substances and high-energy reactions. Predictive analytics enhances OHS in such laboratories through the surveillance of environmental variables such as temperature, humidity, and gas concentrations. In laboratories dealing with flammable solvents, predictive models trained on ventilation data and gas sensor readings can anticipate dangerous vapor accumulations. In high-energy physics setups like particle accelerators, predictive analytics monitors magnetic field intensities and vacuum chamber conditions to prevent equipment failures and radiation leaks. Radiochemistry and radiology labs involve handling radioactive isotopes and high-energy radiation sources. Predictive analytics plays a vital role in radiation safety by analyzing dosimetry data, equipment usage patterns, and shielding conditions. Models predicting radiation exposure levels based on historical usage data help optimize worker rotation schedules, minimizing individual exposure. In radiology labs, predictive analytics can forecast X-ray tube failures by analyzing operational parameters, reducing unplanned downtime and ensuring patient and worker safety.

Predictive analytics has shown potential in water resources research. It can enhance the monitoring of OHS risks in lacustrine, potamic (riverine), and marine zones. In lacustrine zones, predictive models analyze water quality parameters like dissolved oxygen, turbidity, and pH levels can forecast algal blooms, which may pose respiratory risks to workers. In potamic zones, real-time monitoring of river flow velocities and sediment loads supports the safety of hydro-engineering projects. Marine applications include predicting offshore platform risks by analyzing wave height, wind speed, and corrosion data, enhancing safety in oceanographic research and resource extraction. These models can be used to preempt shark attacks and jellyfish attacks by forecasting the times when the probability of the presence of such creatures in a region is the highest. Variables such as weather, climate and other environmental conditions, recent events like storms, hurricanes, earthquakes, tsunamis, landslides (which may cause river blockages), and historical data detailing the frequency and times of attacks can be used to expedite this goal.

In electronics manufacturing, predictive analytics is pivotal in ensuring worker safety amidst the use of hazardous chemicals such as soldering fluxes and cleaning solvents. Predictive models can analyze air quality sensor data and forecast harmful levels of volatile organic compounds (VOCs) to instigate the maintenance staff to undertake appropriate ventilation adjustments. Predictive maintenance systems for robotics and conveyor equipment reduce the risk of mechanical failures. Advanced analytics also aids in mitigating electrostatic discharge risks. These are critical in cleanroom environments where sensitive electronic components are handled.

Predictive analytics has demonstrated its effectiveness in OHS through various global implementations. In the manufacturing sector, General Electric (GE) employs predictive analytics to monitor its turbine and jet engine operations, reducing equipment failure rates by a significant level. Global construction firms use predictive analytics to assess site-specific risks by integrating weather data, workforce behavior, equipment usage patterns, and so on. In the mining sector, Rio Tinto is a British-Australian firm which utilizes predictive analytics to its advantage, Some mining companies monitor vehicle collisions in open-pit mines through driver behavior data and data on environmental conditions. It reduces worker accidents. Algorithms in the logistics industry analyze accident trends and driver behavior data to prioritize safety training for high-risk drivers. This targeted approach reduces costs associated with overgeneralized interventions while improving safety.

## 4. Methodology for Implementing Predictive Analytics

The foundation of predictive analytics lies in robust data acquisition. In India, key data sources include IoT-enabled sensors installed on machinery, wearable devices for workers, safety audit reports, and environmental monitoring systems. Additional data can be gathered from incident logs, maintenance records, government safety compliance databases, and rules, regulations, and legislations. In high-risk sectors such as mining, real-time geospatial and environmental data from satellite imagery and drone surveillance can complement on-ground sensor data. Ensuring data accuracy, consistency, and completeness is necessary for effective predictive analytics. Automated verification techniques, such as anomaly detection algorithms, to identify and rectify erroneous entries. Data preprocessing methods, including normalization, outlier and noise removal, missing value imputation, and feature engineering are essential for creating reliable datasets. Given the diversity of India's industrial sectors, data standardization protocols must be established to harmonize inputs from disparate sources.

Ethical considerations must be one of the concerns at the forefront of a successful data analytics paradigm. Worker privacy and data protection are paramount. Adherence to India's evolving data protection framework, such as the Digital Personal Data Protection Act 2023, must be prioritized. Mechanisms for anonymizing sensitive data, obtaining informed consent from workers, and restricting access to critical safety data ensure ethical compliance and inspires trust among stakeholders.

The selection of algorithms should be guided by the complexity and scale of the data. For predicting machine failures or safety incidents, supervised learning models such as random forests, gradient boosting machines (e.g., XGBoost), and neural networks are effective. Time-series forecasting models, including Long Short-Term Memory (LSTM) networks, are suitable for analyzing sequential data, such as temperature fluctuations or vibration signals. Feature selection strategies must emphasize context-relevant variables, such as equipment age, worker shift duration, and environmental factors. Using pretrained models and techniques like finetuning and transfer learning speeds up the process.

Complementary to deep learning (neural networks), statistical methods such as regression analysis (logistic regression and linear regression), Markov models, and Bayesian inference can be employed for probabilistic risk assessment and trend analysis. These methods can be useful in industries with limited datasets, such as SMEs in the informal sector. Developing models that emphasize interpretability is critical for widespread adoption. Explainable Artificial Intelligence (XAI) techniques like SHAP (SHapley Additive exPlanations) values can be used to explain model predictions. This makes it easier for safety officers to understand and act on insights. It also helps make the AI models and their output trustworthy.

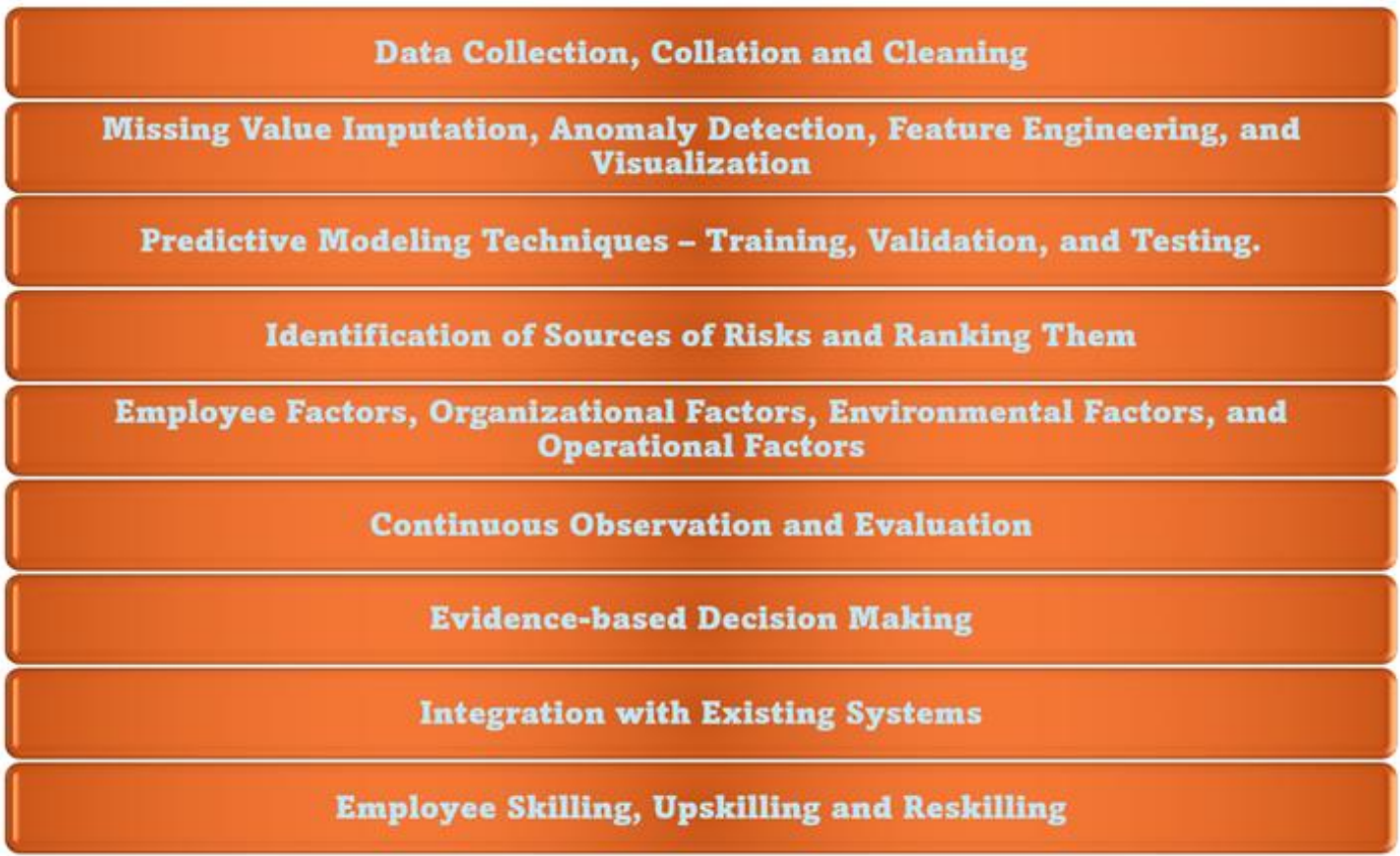

**Figure 1: Summary of Various Concepts and Considerations Associated with the Methodology of Implementing Predictive Analysis Within Occupational Health and Safety in an Organizational Setup**

In relevant settings, predictive analytics platforms must seamlessly integrate with existing enterprise resource planning (ERP) systems, safety management software, and supervisory control and data acquisition (SCADA) systems. API-based integration ensures interoperability and real-time data sharing across platforms. A significant barrier to adoption in India is the lack of technical expertise among safety personnel. Comprehensive training programs in data literacy, model interpretation, and decision-making based on predictive outputs are essential. Partnerships with academic institutions and industry bodies can facilitate capacity building. Implementing predictive analytics in a phased manner, starting with pilot projects in high-risk industries such as construction or chemical manufacturing, allows for iterative refinement of models and systems. Lessons learned from pilot implementations can inform broader rollouts across sectors.

In the context of predictive analytics, cloud platforms provide scalable storage and computational resources. Cloud platforms enable the deployment of predictive analytics at lower upfront costs. Some cloud services providers offer pay-as-you-go services which are beneficial for SMEs and startups. In regions with limited connectivity, edge computing devices can enable the processing of data locally. Edge computing, in such cases, reduces costs associated with predictive analytics and latency, as well as ensures that ongoing monitoring can continue in remote locations such as mining regions and areas with offshore platforms. Open-source machine learning frameworks like Keras, TensorFlow, PyTorch, and Scikit-learn provide pretrained models and inbuilt functions that can potentially reduce software development costs. Open-access safety data repositories can be developed in collaboration with governmental and non-governmental organizations. Such repositories can democratize access to high-quality datasets for effective predictive analytics in OHS in India. Possible government initiatives, such as subsidies for safety technology adoption and tax incentives for deploying predictive analytics systems, can incentivize the implementation of intelligent predictive analytics models. Public-private partnerships (PPP) can help share the financial burden of initial deployments as well as streamlining maintenance.

## 5. OHS: The Case of India

India is a developing country. India boasts of a diverse occupational tapestry. This tapestry constitutes a pastiche of sectors divided under two overarching heads, organized sector (unionized sector with better legal coverage and protections) and unorganized sector. The industrial organization of India faces some unique challenges in maintaining workplace safety and employee wellbeing. Within this fabric, predictive analytics exemplifies immense potential for improving OHS [1]. Predictive analytics can be instrumental in addressing OHS challenges in the Indian context.

India's uniqueness lies in the fact that it is an emerging economy [2] [3], where child labor continues to be practiced clandestinely, where female and child involvement continues to be high in physically demanding and hazardous occupations, where manual cleaning and dredging of sewers, septic tanks, and manholes continues even though a stern mandatory cessation has been imposed on manual scavenging, where males as well as females overwhelmingly participate in informal labor, where precautions like using proper safety gear are frequently not adhered to, where standards regarding polluting substances including poisonous gases released in the workplace exist (for instance, the Indian Government's Ministry of Steel's "SAFETY CODE FOR IRON & STEEL SECTOR" seeks to curb the negative effects of hazards like the "presence of toxic, flammable or asphyxiating gases", especially in "confined spaces".) but with a weak system of implementation, and where a uniform law for OHS is absent and the legal framework in this regard is propped up by disparate, sector-centered laws which causes confusion and leaves many sectors untouched by its ambit. In this regard, predictive analytics can act as a transformational tool in remedying this situation if combined with systemic changes within the domain, especially at the legislative and executive levels.

In January 2022, six people employed at a dyeing and printing mill in Gujarat in India died and about twenty of them had to be hospitalized after breathing in poisonous fumes which emanated from an illegal discarding of excess hazardous chemicals in a rivulet [4]. On April 18, 2023, four migrant employees of a decrepit rice mill were killed and twenty of them were injured after the mill, in which the laborers were sleeping, collapsed in Haryana. On April 26 the same year, three sanitation workers in India died because a lack of safety gear caused them to inhale toxic fumes as they worked inside a septic tank in West Bengal [5]. On the morning of October 27, 2024, two workers died and seven had to be rendered to hospital care after they breathed in poisonous fumes working in proximity to a textile factory [6]. Manual sewerage cleaning has been termed the "world's worst job" and the one of the most dangerous and fatal occupations in the country [7]. After the Bhopal Gas Tragedy in 1984, which occurred forty years ago, poisonous gas inhalation deaths and hospitalizations continue to happen unabated in India. Multiple mining mishaps, including the 1965 (Dhori colliery disaster) and 1975 (Chasnala mining disaster) coal mine incidents, both in Jharkhand [8], worker-centered precautions with regards to mining, especially coal mining, continue to be overlooked. Many other occupations face similar situations in India. These statistics suggest that India's OHS scenario needs a radical overhaul.

India has over 90% of workers employed in unorganized industries. This leads to a lack of OHS coverage and regulations for informal workers, difficulty in implementing and enforcing safety standards, and limited access to occupational health services for informal workers [9, 10]. For India's vast informal sector, predictive models can be developed to identify vulnerable worker groups in unregulated environments, using data sources like mobile-based surveys and public health records. This can help create targeted, cost-effective safety initiatives and prioritize coverage areas until formal regulations are enacted. Predictive analytics can also aid in designing localized safety interventions using region-specific data. These interventions can be promoted through mobile platforms to reach informal workers who lack conventional OHS coverage. Cellular phone penetration has reached high levels in India and collection of data and dissemination of such information are potential advantages that can be reaped off it.

The diversity of India's workforce, spanning from agriculture to high-tech industries, requires predictive analytics to offer flexible and sector-specific insights. Machine learning algorithms can model the likelihood of occupational hazards in various industries. This will enable policymakers to tailor training and preventive measures to the specific risks present in each occupation. In high-risk sectors like agriculture and construction, predictive models could use historical accident data to guide interventions and determine industry-specific training needs. Such interventions can cater to a wide variety of work environments.

India's rapid industrialization and urbanization bring new occupational hazards and strain existing OHS infrastructure. There is a potentially increased exposure to occupational hazards as new industries emerge. This puts strain on existing OHS infrastructure and resources and demonstrates the need for frequent updates to safety regulations to keep pace with industrial growth. Predictive analytics can be used to forecast emerging risks associated with new industries and highlight potential OHS risks before they escalate. This approach supports timely regulatory updates. Predictive analytics can also model the optimal allocation of limited OHS resources. It helps maximize the impact of interventions in high-risk industrial and urban areas.

Socioeconomic challenges in India are vast. They comprise prevalence of child labour in certain sectors, the large migrant worker population with poor living conditions, and high poverty rates leading to workers prioritizing income over safety, among other struggles Indias workers face. These, in relation to OHS, can also be tackled by predictive models. Predictive models can identify areas where workers are most susceptible to compromising safety for income. This can be carried out by integrating socioeconomic data with health and labour information. In this manner, predictive models can mark vulnerable populations. They can recommend targeted policies and resource allocation measures to improve workplace safety for these groups. For India's large migrant workforce, predictive analytics can track health patterns across regions and help provide appropriate OHS support based on real-time needs and living conditions.

The limitations associated with India's OHS infrastructure, such as the shortage of trained OHS professionals, lack of abundant data collection, and a deficiency of centralized oversight by an overarching regulatory body, can also be mitigated by predictive analytics [11, 12]. Predictive models can help prioritize inspections, identify regions and industries with high occupational risk, and direct the limited workforce to areas where interventions are most needed. Predictive analytics can recommend courses and training programs for existing professionals and spheres where the best OHS professionals are most likely to be found so that they can assist in policy formulation. For example, if a certain business administration program has shown consistently excellent outcomes with regards to OHS professionals, policy formulation advisors can be chosen from among the graduates of that program. Predictive analytics recommendations can facilitate centralized data collection and reporting systems. These exercises enable the establishment of a unified OHS knowledge base that can be continually updated to reflect current workplace hazards and needs.

Cultural and awareness-related challenges, including low OHS awareness, covert adherence to hazardous cultural practices like manual scavenging, and apathy towards OHS among some stakeholders, may be addressed with predictive models that map awareness gaps and vulnerable communities. Using culture-specific information and data about previous policies and interventions regarding OHS, and in the situations in which those policies and interventions were formulated and implemented, the model can suggest the best administrative measures that need to be taken for specific cultures, especially with regards to their beliefs and the specific hazardous practices they follow or dangerous beliefs they adhere to. Anthropological and sociological approaches come handy in such cases. Targeted awareness campaigns will ensure that outreach efforts are evidence-based and culturally sensitive. Predictive analytics may be employed to identify the likelihood of unsafe practices within specific communities and sectors. Such an approach facilitates directed interventions to shift these norms toward safer practices.

India's workforce faces a dual disease burden in their workplace. There are traditional occupational illnesses like silicosis, chronic obstructive pulmonary disease, chemical dermatitis, and so on. Accompanying them are modern lifestyle-related conditions like diabetes and cardiac conditions. The prevalence of such a disease burden may benefit from the use of predictive analytics. Predictive models can track the prevalence of both categories of diseases and identify their common predictors. This helps in prioritizing disease prevention strategies tailored to different workforce segments. Outlining the progression of lifestyle diseases can aid OHS initiatives to incorporate preventive health programs within workplaces.

In India, the existence of a multiplicity of statutory controls within the sphere of OHS leads to misunderstandings and muddle ups. The inadequate enforcement of existing legislations and the lack of OHS coverage to all sectors, particularly unorganized ones, are further concerns. India's workforce spans a wide range of industries and occupations, from traditional agriculture and small industries to modern IT services and medical facilities. This diversity makes it challenging to develop comprehensive OHS policies applicable to all sectors, provide specialized safety training across varied occupations, and address industry-specific hazards effectively. In relation to this, predictive analytics may be used to recommend the points and sections which can be used to construct an all-embracing legal code for OHS in India which especially protects workers in the unorganized sector and endeavors to formalize as many workers as possible.

## 6. Barriers to the Adoption of Predictive Analytics in OHS in India

The adoption of predictive analytics in OHS in the Indian context faces many barriers that stem from systemic, technological, and socio-economic constraints, as well as from the unique regulatory and cultural milieu of the country. The lack of comprehensive and standardized safety data across industries in India is a significant impediment. Many organizations, particularly in the informal and SME sectors, do not maintain detailed records of workplace incidents, equipment performance, and environmental conditions. This results in sparse OHS datasets which are insufficient for building robust predictive models. Some sectors also lack sector-specific data.

The quality of available data is frequently compromised by inaccuracies, inconsistencies, and missing values. Manual record-keeping practices and poorly maintained digital logs lead to unreliability in datasets. The security of sensitive worker data as well as organizational data must be ensured, especially because of the growing digitization of OHS records. Digitization heightens the risk of data breaches and unauthorized access has increased. There are potentially many industries in India which lack robust cybersecurity measures. This can cause predictive analytics systems in OHS in India to become vulnerable to exploitation. Standardized cybersecurity protocols lower such risks.

A number of industrial zones in India, particularly in rural and semi-urban areas, are plagued by inadequate connectivity and outdated equipment. This poses obstructions to the access to datasets to industries in such regions as well as to the implementation of cloud computing-based models. IoT sensors needed for real-time data collection fail to function reliably in regions with intermittent power supply. Many Indian industries, especially those in traditional sectors such as textiles or agriculture-based manufacturing, are not technologically equipped to adopt predictive analytics. Legacy systems and traditional (sometimes obsolete) machinery such as handlooms do not support the integration of modern data acquisition tools and predictive algorithms.

A strong reliance on reactive approaches to OHS often results in resistance to adopting novel technologies which are proactive in nature. Workers and managers can be skeptical about the efficacy of predictive analytics, the perception that predictive analytics is complex and intrusive, and the costs related to its adoption. Implementing and utilizing predictive analytics requires a workforce with some expertise in data science, machine learning, and safety management. India faces a skill gap in these areas. There is limited availability of trained personnel at both technical and managerial levels. Industries that can afford predictive analytics solutions may struggle to recruit and train workers capable of interpreting and acting on predictive insights. Skilling, upskilling and reskilling the workers is needed for the implementation of predictive analytics in OHS in India.

However, overreliance on predictive analytics in OHS can also cause obstacles. There must be qualified human experts and employees who should be able to validate the outcomes of an AI model tasked with predictive analytics in OHS. While it is a serious requirement for such models to be designed to be as precise and as effective in their decisions as possible, the possibility that they can make errors must not be overlooked.

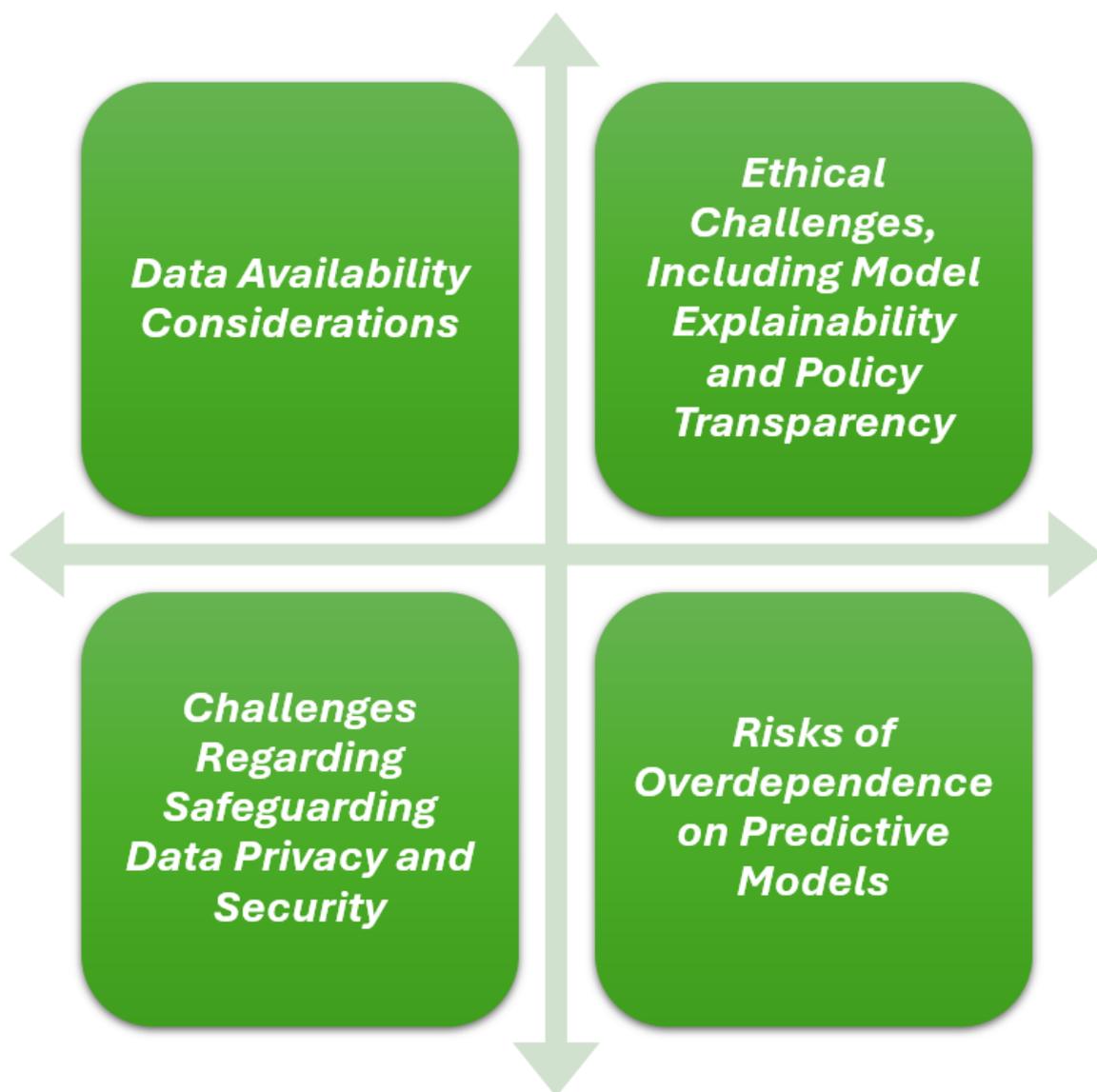

Figure 2: Challenges Associated with Predictive Analysis in Occupational Health and Safety

India has made strides in workplace safety regulations, such as the Occupational Safety, Health, and Working Conditions Code, 2020, the Code on Social Security, 2020, the Code on Wages, 2019, and the

Industrial Relations Code, 2020. Nevertheless, most of these laws and legislations continue to lack provisions for the adoption of advanced technologies like predictive analytics. There are no specific guidelines for the ethical use of predictive models in OHS. This leads to organizations having to traverse this space independently, on their own. Inconsistent enforcement of existing safety standards undermines the urgency of adopting innovative solutions. In India, the awareness of digital rights and data privacy is still evolving. Therefore, associated ethical concerns can become a significant barrier to the acceptance of predictive analytics in OHS in India. They must be remedied to ensure the seamless incorporation of predictive analytics in the OHS setup of India. Necessary provisions from international privacy protection laws such as Europe's GDPR can be borrowed or adapted to enhance India's revamped privacy protection framework [13].

## 7. Policy Recommendations and Future Directions

The policy framework in India must be such which incentivizes adoption of predictive analytics in OHS. Both central as well as state governments should introduce a well-thought-out combination of tax benefits, subsidies, and low-interest loans to encourage the adoption of predictive analytics in safety management. Such financial incentives would reduce the cost burden of adopting such technologies on SMEs. Regulatory frameworks should be updated to include mandatory provisions for predictive analytics in high-risk sectors such as mining, construction, and chemicals. Periodic digital risk assessments operated by predictive tools may be made a compliance requirement.

A national-level open data repositories, where anonymized safety and operational data from industries is aggregated, must be established. Such a repository will provide valuable data for model training and innovation. Publicly accessible datasets would also spur academic and corporate research collaborations in machine learning and predictive analytics as applicable to OHS in India. Standardization of predictive analytics tools and methodologies through certification bodies would ensure reliability and interoperability across sectors. Industries adopting certified tools could gain competitive advantages or compliance credits in tenders and government contracts.

PPPs are the vehicles that can drive innovation in predictive analytics by utilizing the expertise of academic institutions, government research bodies, and private technology firms. Collaborative projects should focus on domain-specific solutions, such as real-time hazard detection in nuclear safety and ergonomic risk prediction in electronics manufacturing. Establishing cross-sector platforms for knowledge exchange will enable industries to learn from global best practices. For example, forums facilitated by the Ministry of Labor and Employment should bring together safety professionals, data scientists, industry representatives, and policymakers to discuss challenges and innovations. PPPs will also facilitate co-funding arrangements to reduce the financial burden on industries. State governments may partner with technology providers to subsidize the cost of IoT and other safety infrastructure in industrial clusters and Special Economic Zones (SEZs).

Ongoing investments continue to occur in robust IT infrastructure, such as high-speed internet in industrial zones, IoT-enabled machinery, and edge computing devices, in India. Such capital expenditure must be government's priority to set India on a path of becoming a manufacturing superpower and a hub of entrepreneurial growth and Foreign Direct Investment (FDI).

Government programs like the Digital India initiative can be expanded to include safety technology in industrial applications. Comprehensive training programs in predictive analytics, machine learning, and safety data interpretation may be incorporated into existing skill development schemes like the Pradhan Mantri Kaushal Vikas Yojana (PMKVY). Custom curricula tailored to safety professionals may bridge the skill gap and enhance workforce readiness. Collaborations with global leaders in predictive analytics will facilitate technology transfer. This will empower Indian industries and enable them to access state-of-the-art predictive analytics solutions in OHS. Partnering with multinational firms for on-site demonstrations and pilot projects will also build confidence in technology adoption.

Future research should focus on developing predictive models tailored to the unique needs of different industries in India. AI models addressing heat stress in agriculture, toxic gas exposure in petroleum refineries, and repetitive strain injuries in garment manufacturing can render highly targeted solutions. The integration of machine learning with domain-specific simulations such as finite element analysis in structural engineering and real-time physics-based modeling must be explored to enhance the precision of predictive systems in OHS.

Worker behavior and psychological data must be incorporated into predictive models in order to create new ways to engender proactive safety measures. A large proportion of Indian middle-class employees face work-associated stress, long commutation times, high costs of rentals and transportation, and variegated household issues. Statistics related to this fact can be incorporated into AI models for predictive analytics for OHS. Fatigue detection using wearables and its integration into risk prediction frameworks will reduce incidents caused by human error. Research into ethical AI practices and bias mitigation in predictive analytics will also be needed to ensure fairness across diverse worker demographics, given India's heterogeneous labor force. Investigating the transferability of predictive models across sectors, such as using insights from nuclear safety to improve radiological lab protocols, must be studied for innovation and facility.

## 8. Conclusion

Realizing the transformative potential of predictive analytics in OHS in Indian environment requires concerted efforts from relevant stakeholders. Administrators must establish a conducive regulatory framework, industries must embrace technology and innovation, and academic and research institutions must drive advancements in predictive modeling tailored to India's unique socio-economic and industrial context. The time is ripe for public-private collaborations to fund infrastructure development, support workforce training, and ensure ethical and scalable implementations. By fostering such synergies, India can overcome challenges related to data availability, technological readiness, and workforce adaptability.

Techniques based on Artificial Intelligence (AI) and its subfields like machine learning, especially deep learning, which includes Large Language Models (LLMs), have revolutionized many fields in the favor of humans, drastically easing routine processes and solving hefty problems [14 - 19]. The field of OHS must also revel and take advantage of the opportunity presented by AI. Predictive analytics is one of the core features of machine learning models. Looking ahead, the future of OHS in India lies in embedding predictive analytics as a cornerstone of workplace safety. India shows promise to create safer and more resilient work environments by way of implementing innovative approaches which form a nuanced amalgam of cutting-edge technology with traditional safety protocols.

An evidence-based approach like predictive analytics is convivial for data-driven risk assessment, optimization of resources, and prevention of safety-related incidents. It results in improved safety outcomes and sustained success. However, while implementing predictive analytics, it is crucial to address challenges like data availability, potential overdependence on predictive analytics models, model explainability, and ethical concerns like data privacy and transparency when integrating predictive models into safety strategies. Collaboration between various stakeholders in OHS, including workers (both formal and informal), representatives of labor unions, managers, supervisors, policymakers, worker and general welfare machinery, healthcare providers, researchers and statisticians, technologists and data scientists, advocacy groups, and grassroots, social, non-profit and non-governmental organizations is needed for bringing about a paradigm shift in India's OHS policy. Further research and development can help overcome extant bottlenecks, reduce implementation costs for various OHS interventions at sub-organizational, organizational, supra-organizational (bureaucratic public policymaking) levels, and help design, develop and execute better predictive analytics models.